\documentstyle[prb,aps,epsfig]{revtex}          

\begin{document}

\title{First-principles study of lattice instabilities in the ferromagnetic 
martensite Ni$_2$MnGa }

\author{Claudia Bungaro$^{*}$ and K. M.~Rabe}

\address{Department of Physics and Astronomy, Rutgers University, 
Piscataway, NJ 08854-8019, USA.}

\author{A.~Dal~Corso}

\address{Scuola Internazionale Superiore di Studi Avanzati (SISSA), 
Via Beirut 2/4, 34014 Trieste, Italy.}

\maketitle

\begin{abstract}
The phonon dispersion relations and elastic constants for
ferromagnetic Ni$_2$MnGa in the cubic and tetragonally distorted
Heusler structures are computed using density-functional and
density-functional perturbation theory within the spin-polarized
generalized-gradient approximation.  For $0.9<c/a<1.06$, the TA$_2$
tranverse acoustic branch along $[110]$ and symmetry-related
directions displays a dynamical instability at a wavevector that
depends on $c/a$.  Through examination of the Fermi-surface nesting
and electron-phonon coupling, this is identified as a Kohn anomaly.
In the parent cubic phase the computed tetragonal shear elastic
constant, C$^\prime$=(C$_{11}-$C$_{12}$)/2, is close to zero,
indicating a marginal elastic instability towards a uniform tetragonal
distortion.  We conclude that the cubic Heusler structure is unstable
against a family of energy-lowering distortions produced by the
coupling between a uniform tetragonal distortion and the corresponding
$[110]$ modulation.  The computed relation between the $c/a$ ratio and
the modulation wavevector is in excellent agreement with structural
data on the premartensitic ($c/a$ = 1) and martensitic ($c/a$ = 0.94)
phases of Ni$_2$MnGa.

\end{abstract}
\pacs{PACS:  }

\twocolumn
\section{INTRODUCTION}

Ferromagnetic shape-memory alloys displaying large magnetic-field-induced 
strain have recently emerged as a new class of active materials,
very promising for actuator and sensor applications.
The largest known magnetostrain effects have been observed 
in Ni$_2$MnGa-based Heusler alloys, where up to 6\% and 9.5\%  strains have 
been induced by a magnetic field less than 1 Tesla.~\cite{MSMA00,MSMA02}

The magnetic shape-memory behavior is closely linked to the occurrence
of a martensitic transformation in conjuction with a strong
magnetocrystalline anisotropy of the low-temperature martensitic
phase. For a deeper understanding of the magnetic shape-memory
mechanism, a microscopic explanation for the origin of the martensitic
transformation and the magnetocrystalline anisotropy is clearly
needed. Towards this goal, we have used ab initio techniques to
investigate the origin of the martensitic transformation.

Ni$_2$MnGa is the most intensively studied of the relatively few known
ferromagnetic shape-memory materials. A number of thermal and
stress-induced martensitic transformations have been observed in
Ni$_2$MnGa-based alloys, and both transition temperatures and
crystallographic structures are quite sensitive to alloy composition.

In the stoichiometric alloy, Ni$_2$MnGa, which is ferromagnetic below
T$_{\rm C}\approx$ 380 K, two thermally-induced phase transitions have
been observed.  From the high-temperature cubic Heusler structure, a
premartensitic phase transformation to a modulated cubic structure
occurs below T$_{\rm PM}\approx$ 260 K,~\cite{Zheludev95,Zheludev96}
followed by a martensitic transformation to a modulated tetragonal
structure below T$_{\rm M}\approx$ 220 K.~\cite{Webster84} The low
temperature martensitic phase has a tetragonal structure ($c/a$=0.94,
$a$=5.90$\AA$) with a superimposed incommensurate modulation along the
[110] direction, consisting of a shuffling of (110) planes in the
[$\overline 1$10] direction with a periodicity of almost 5 interplanar
distances, which corresponds to a wave vector {\bf q}$_{\rm
M}\approx{2 \pi \over a}$(0.43, 0.43, 0)
(Refs. \onlinecite{Zheludev96,Martynov92}).  The low-temperature
martensitic phase has a high magnetocrystalline
anisotropy,~\cite{magAnis} making it useful for applications.  At a
phenomenological level, the coupling between strain, modulation, and
magnetization has recently been described in a Landau-theory
framework.~\cite{phen}

Some clues to the microscopic origin of the martensitic transitions are
provided by the softening of particular
phonons and elastic constants. Inelastic neutron scattering experiments
on the high-temperature phase found a
significant, though incomplete, softening in the TA$_2$ phonon branch
along the [110] direction, at a wave vector 
{\bf q}$_{\rm PM}\approx{2 \pi \over a}$(0.33, 0.33, 0) 
(Ref. \onlinecite{Zheludev95}).  
The phonon softening has been found to correspond to a premartensitic
phase occurring between the high-temperature cubic and low-temperature 
martensitic structures. The premartensitic phase has a  
cubic structure with a superimposed [110]-transverse shuffling modulation, 
analogous to the modulation of the martensitic phase but with a different 
periodicity of almost 6 interplanar distances.~\cite{Zheludev95}
While the PM phase is anticipated by a precursor phonon softening at  
{\bf q}$_{\rm PM}$, no phonon softening at {\bf q}$_{\rm M}$ 
has been observed above T$_{\rm M}$.

The aim of this study is to provide a unified explanation
for the microscopic origin of the rich
variety of phase transitions and modulated structures 
occurring in this ferromagnetic shape-memory alloy.
To this end we have performed a first-principles study of
the phonon dispersions and 
lattice instabilities of ferromagnetic Ni$_2$MnGa and of their 
dependence upon uniform tetragonal strain. 
In our calculations we use the spin-polarized generalized-gradient 
approximation ($\sigma$-GGA)
recently developed within the density-functional perturbation theory 
(DFPT) formalism.~\cite{DalCorso00}  This
approach allows us to obtain an accurate description of the
structural, magnetic, and vibrational properties of Ni$_2$MnGa.

The paper is organized as follows. The details of the computational method 
are given in section II. In section III we present the results for the 
cubic Heusler structure. The crystal structure and magnetization are discussed
in Section III.A. Section III.B is devoted to the 
phonon dispersion, with particular attention to the
phonon anomaly and related dynamical instability.
The origin of the phonon anomaly is discussed in Section III.C. 
The computed elastic constants are presented and discussed in Section III.D.
In Section IV we investigate the dependence of 
the phonon dispersions and lattice instabilities on uniform tetragonal strain.
In Section V the main results are summarized and the 
microscopic origin of the observed phase transitions explained.

\section{COMPUTATIONAL METHOD}
Our calculations have been performed within the framework of
density-functional and density-functional-perturbation theory 
(DFT and DFPT).
In particular, the vibrational properties
have been computed using a recent implementation of ultrasoft pseudopotentials
into DFPT.~\cite{DFPTUS} 
We used ultrasoft
pseudopotentials~\cite{ultrasoft} for Ni and Mn, freezing the 3$s$ and
3$p$ core electrons and treating the 3$d$ and 4$s$
states as valence levels.~\cite{Nips,Mnps} The nonlinear core
correction is used to account for the overlap between the core and the
valence charges.~\cite{NLCC}  For the Ga atom we used a norm-conserving
pseudopotential, which also includes a nonlinear core correction,
treating the 4$s$ and 4$p$ states as valence levels.  To describe
the effects of exchange and correlation, we used the
Perdew-Burke-Ernzerhof\cite{PBE} functional
with the spin-polarized generalized gradient correction ($\sigma$-GGA), 
recently implemented within the DFPT formalism in the PWscf 
code.~\cite{DalCorso00} For comparison, 
we have performed calculations within the local spin density
approximation (LSDA) using the Perdew-Zunger parameterization of the
exchange and correlation energy.  The plane-wave basis set had
a kinetic-energy cutoff of 25 Ry.  The augmentation charges,
required by the use of ultrasoft pseudopotentials, were expanded 
with an energy cutoff of 450 Ry.  
The Brillouin zone integration was
performed using the smearing technique,~\cite{sme-Meth} suitable for
metallic systems.  The structural properties and most of the phonon
frequencies are well converged using a 
first-order smearing function with a
smearing parameter $\sigma$=0.03 Ry and an fcc (6, 6, 6) Monkhorst-Pack
grid,~\cite{kmp} yielding 28 {\bf k}-points in the
irreducible wedge of the Brillouin zone (IBZ) for the cubic structure.
To obtain an accurate description of the anomalous
TA$_2$ phonon branch, a smaller smearing parameter is necessary and, 
consequently, a finer k-point sampling. 
An accuracy to within a few cm$^{-1}$ has been obtained using 
$\sigma$=0.01 Ry and an fcc (10, 10, 10) k-point grid, 
yielding 110  {\bf k}-points in the IBZ, for the cubic structure.
For the tetragonal structure, an (8, 8, 8) k-point grid 
yielding 144  {\bf k}-points in the IBZ of the face-centered orthorhombic
unit cell was used.

To compute the full phonon dispersions of the cubic structure, 
we computed the interatomic force constants by Fourier transformation
of the dynamical matrices computed on a (6, 6, 6) {\bf q}-point
grid in the fcc BZ.
The phonon dispersions along the [110] ([011]) direction were obtained 
by interpolating the dynamical matrices computed on a finer mesh of 
24 {\bf q}-points between $\Gamma$ and the shortest reciprocal-space 
vector parallel to the [110] ([011]) direction. 
An even denser mesh of {\bf q}-points has been used in the proximity 
of the anomaly for a more accurate interpolation of the TA$_2$ branch.

\section{Cubic N\lowercase{i}$_2$M\lowercase{n}G\lowercase{a}}

\begin{figure}
 \centerline{\epsfig{figure=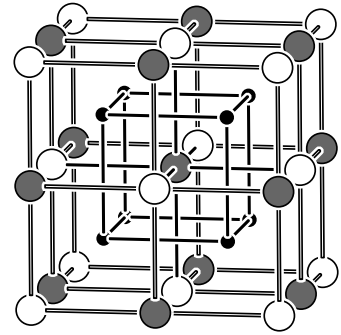,height=5cm,angle=-90}}
\caption{The fcc L2$_1$ Heusler structure of Ni$_2$MnGa. The small
circles, large open circles and large filled circles represent
Ni, Mn and Ga, respectively.}
\label{fig:cell}
\end{figure}

In this section, we report the results of calculations for Ni$_2$MnGa 
in the fcc L2$_1$ Heusler structure 
(see figure~\ref{fig:cell}).

\subsection{Crystal structure and magnetization}

In table~\ref{tab:structure} are given the minimum-energy lattice
parameter, $a_0$, the bulk modulus, B$_0$, and the magnetic moment per
unit cell, $\mu_0$, computed within the spin-polarized
$\sigma$-GGA. Note that to obtain convergence to within 0.01$\mu_B$,
$\mu_0$ has been computed on a (10, 10, 10) k-point grid with
$\sigma$=0.01 Ry.  These results agree well with those of all-electron
calculations (FLAPW) performed using the $\sigma$-GGA.~\cite{Ayuela99}
The theoretical results are also in very good agreement with the
experimental data; the theoretical lattice parameter obtained with
$\sigma$-GGA is equal, to within theoretical precision, to the
experimental value.  For comparison, the LSDA results are included in
table~\ref{tab:structure}.  The LSDA lattice parameter is 2.5\%
smaller than the experimental value.  This underestimate of the
lattice parameter is correlated with an increase in the bulk modulus
to 38\% higher than the experimental value.

\subsection {Phonon dispersion relation}

\begin{figure}
 \centerline{\epsfig{figure=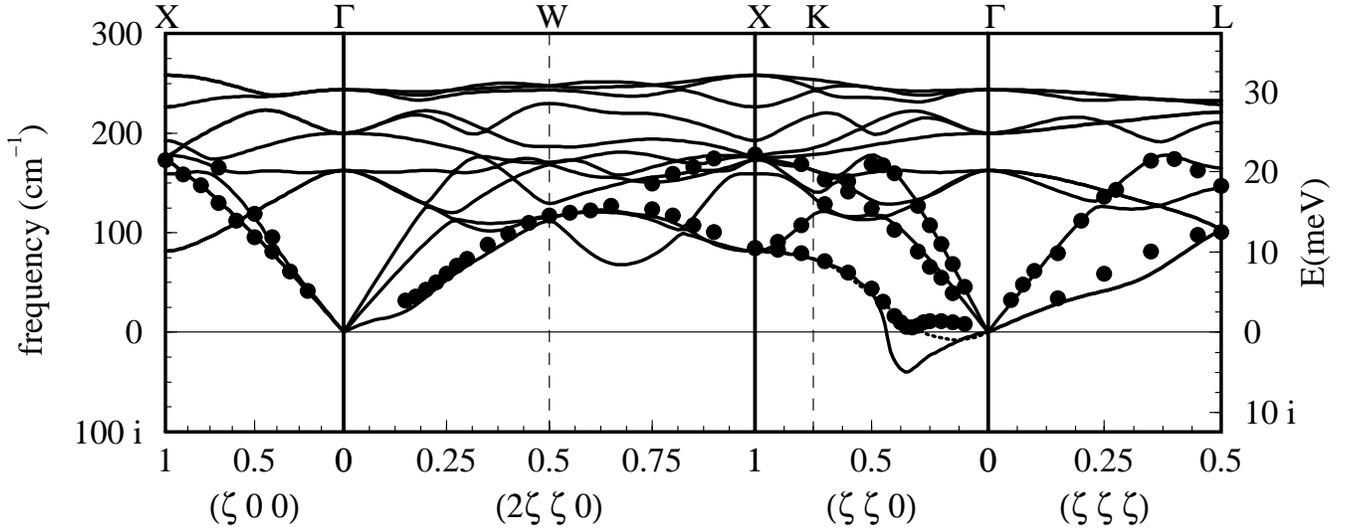,height=7.5cm}}
\caption{Full phonon dispersion of ferromagnetic
Ni$_2$MnGa in the fcc Heusler structure, 
along high symmetry lines of the fcc BZ.
Solid lines are fully converged ab initio calculations obtained using 
an effective electronic temperature $\sigma$=0.01Ry. The dashed line 
is the dispersion obtained using a higher effective electronic temperature, 
$\sigma$=0.03Ry, showing the dependence 
of the anomaly in the lowest acoustic branch along the $(\zeta \zeta 0)$ direction 
on the effective electronic temperature. The circles indicate
the neutron scattering data from 
Refs.~[\protect\onlinecite{Zheludev95-2,Zheludev96}].
The imaginary values of phonon frequencies are plotted along the 
negative frequency axis. The wave-vector coordinate $\zeta$ is 
in units of ${2 \pi \over a}$.}
\label{fig:phcubic}
\end{figure}

The phonon dispersions of the ferromagnetic
cubic structure ($a_0$ = 11.03 a.u.) have been computed, using $\sigma$-GGA,
for $\vec q$ along high-symmetry lines in the first Brillouin zone, as shown
in figure~\ref{fig:phcubic}. The solid
lines indicate the computed phonon dispersion converged to
within a few wave numbers. The theoretical phonon dispersion curves are 
in excellent agreement with the available inelastic
neutron scattering data.~\cite{Zheludev96,Zheludev95-2} 

To evaluate the accuracy of the theoretical method,
the phonon frequencies computed 
using different approximations for the exchange and correlation energy,
at the zone boundary X point, 
are given in table~\ref{tab:Xfreq}. 
Since the
experimental frequency of the anomalous ($\zeta \zeta 0$)-acoustic branch 
is not strongly temperature dependent at this q-point, it is
a good reference for evaluating the accuracy of the 
theoretical method.  The experimental frequency for the transverse
acoustic mode X$_5^\prime$ is in very good agreement with the
phonon frequency computed using the $\sigma$-GGA, with the computed
value being about 6\% softer.  For comparison, we 
have also done the computation with LSDA. Calculations for the cubic
structure with the LSDA lattice constant yield frequencies about 
10\% harder than those computed using the
$\sigma$-GGA, except for the lowest X$_5^\prime$ mode which is 7\%
softer, increasing the discrepancy with experiment. 
In addition, the ordering of the modes in LSDA is different.
Specifically, the X$_4^\prime$ mode is softer than the nearby X$_5^\prime$ and
X$_5$ modes.  
We have also computed the phonon frequencies
within the LSDA but fixing the lattice parameter to the equilibrium value
computed in the $\sigma$-GGA. The result is that 
all the modes soften, so that the frequencies are about 10\% smaller 
than with $\sigma$-GGA. This
softening results in an even greater discrepancy with the
experimental X$_5^\prime$ mode (20\%).  We conclude that the
better accuracy of $\sigma$-GGA is not merely an effect
of the more accurate value of the equilibrium lattice parameter. 
All calculations reported below were performed with $\sigma$-GGA.

The most striking feature of the phonon dispersion relation in 
figure~\ref{fig:phcubic} is the anomalous dip in the lowest branch
of the transverse acoustic modes (TA$_2$) along $[$110$]$. The minimum at the 
incommensurate wave vector 
{\bf q}$_0$=${2\pi\over a}$($\zeta_0$, $\zeta_0$, 0), with $\zeta_0=0.34$, 
is at imaginary frequency.  Thus, the crystal is dynamically unstable 
to the lattice distortion corresponding to the eigenvector of this mode, 
which consists
of a nearly rigid displacement of the (110) atomic planes along the 
[1$\overline 1$0] 
direction with a modulation period of slightly less than 6 interplanar 
distances along the [110] direction. This energy-lowering
distortion can be specified by an amplitude, {\bf u}, and a phase, $\phi$,
\begin{equation}
{\bf u}_m={\bf u}~cos(m \zeta_0 \pi + \phi),
\end{equation} 
where {\bf u}$_m$ is the displacement of the $m^{th}$ (110) 
atomic plane along the [1$\overline 1$0] direction.
Due to the cubic symmetry, there are 12 equivalent anomalies along
the $<$110$>$ directions which give rise to six different but 
crystallografically equivalent lattice modulations ${\bf u}_m$.

\begin{figure}
 \centerline{\psfig{figure=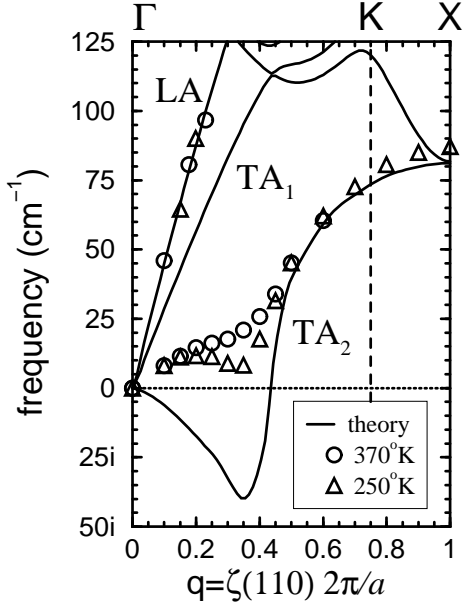,width=6cm}}
\caption{Partial phonon dispersion of Ni$_2$MnGa in the fcc Heusler structure,
along the $\Gamma$-K-X line in the [110] direction. The theoretical
data are the same as in figure~\protect{\ref{fig:phcubic}}. The experimental 
data taken at 250 K and 270 K are shown for comparison. }
\label{phcubicExp}
\end{figure}

The phonon anomaly is shown in more detail in figure~\ref{phcubicExp},
where the theoretical dispersion is compared with inelastic neutron
scattering data taken at two different temperatures: T=370 K (squares)
and T=250 K (triangles).~\cite{Zheludev95} The experimental data
display an anomaly in the TA$_2$ branch that corresponds to the
anomaly predicted by our theoretical dispersion. The wave vector of
the experimentally-observed anomaly is $\zeta_{\rm PM} \approx
0.33$,~\cite{Zheludev95} in excellent agreement with the theoretical
value.  For those modes that do not have a strong dependence upon
temperature, such as the LA branch and the TA$_2$ modes away from the
anomaly ($\zeta\geq$ 0.45), the agreement between theory and
experiment is very good.  To compare theory and experiment for modes
that are strongly temperature dependent, we need to extrapolate the
experimental values to T = 0 K.  In accordance with the soft mode
theory, $(\hbar\omega)^2$ is experimentally observed to decrease
linearly with temperature above T$_{\rm PM}$=260 K. Extrapolation to 
T = 0 K gives an imaginary frequency of $30i$ cm$^{-1}$, which is in
reasonable agreement with the computed value.

\begin{figure}
\vbox{
\centerline{
 \psfig{figure=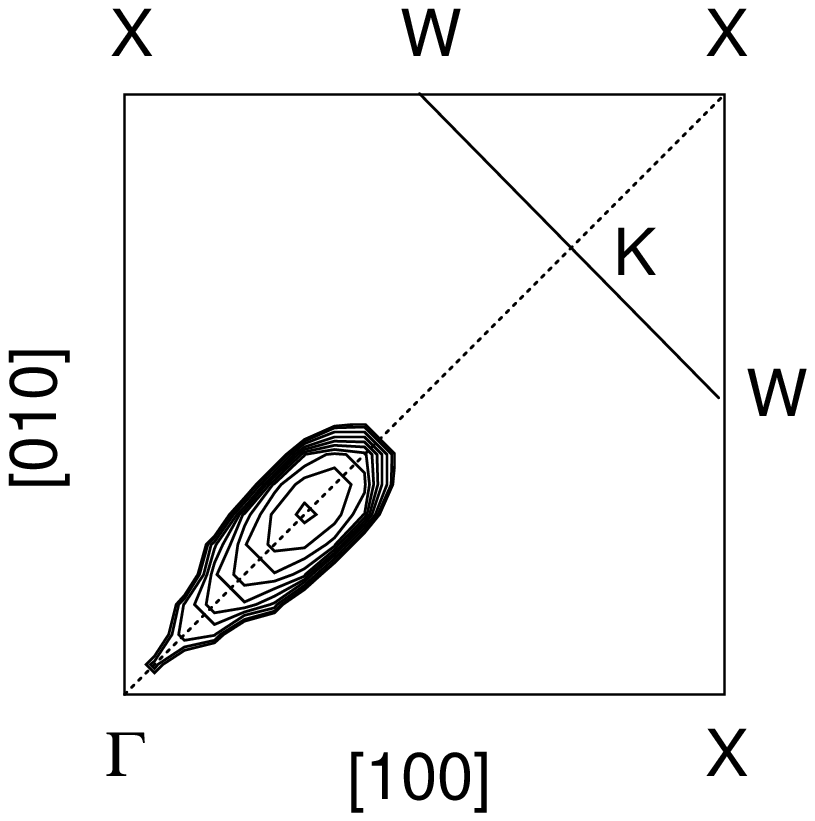,height=4.3cm,angle=0}
}
\vspace{0.5cm}
\centerline{
 \psfig{figure=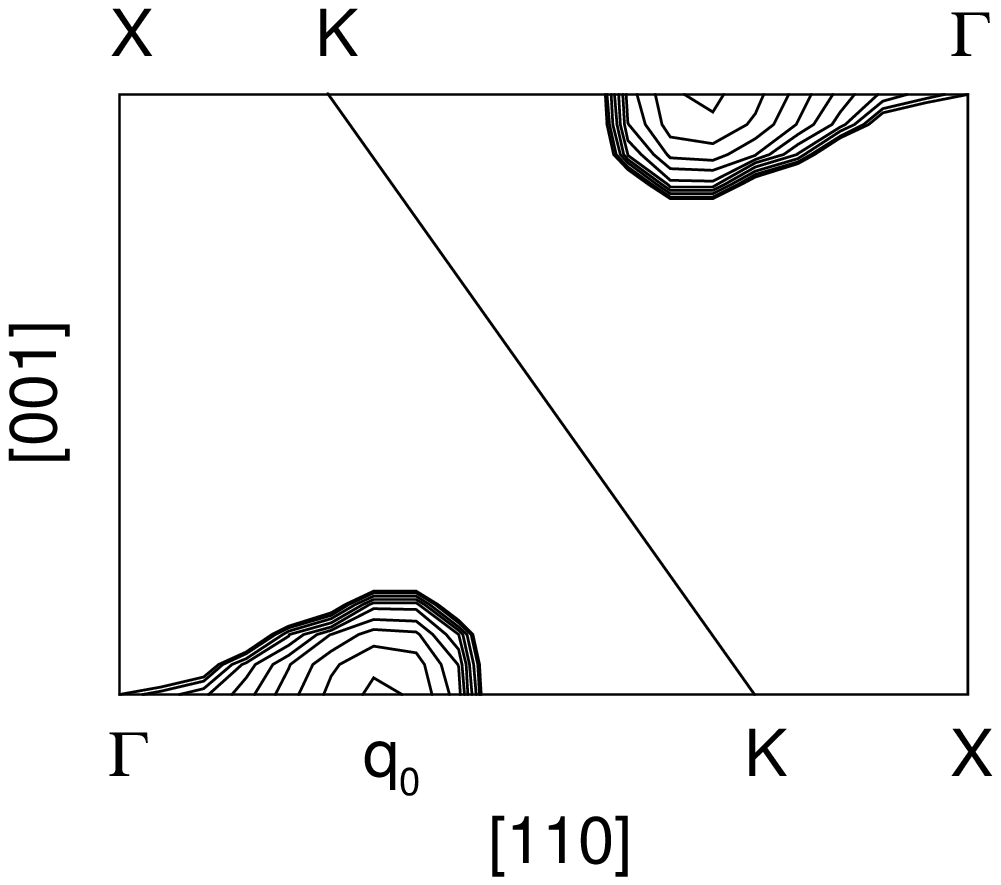,height=4.3cm,angle=0}
\vspace{0.2cm}
}
}
\caption{Contour plots of the imaginary phonon frequencies in the
(001) plane (a) and (1$\overline{1}$0) plane (b) of the fcc BZ. 
These contour plots show the region in q-space
where the crystal is dynamically unstable. This occurs in a narrow valley
situated along the [110] direction with its minimum at the critical wave 
vector {\bf q}$_0$=(0.34, 0.34, 0).}
\label{fig:phcontour}
\end{figure}

Examination of the phonon dispersion throughout the entire BZ shows that 
imaginary frequencies leading to dynamical instabilities occur only in
a very localized region in q-space. This is in agreement with the experimental
dispersion measured in the direction perpendicular to {\bf q}$_0$,
{\bf q}={\bf q}$_0$+($\zeta,-\zeta,0$).~\cite{Zheludev96}
The contour plot in figure~\ref{fig:phcontour} 
shows that the lattice instability 
is confined to a ``drop-shaped'' small region in reciprocal space, located 
along the [110] direction and centered at the critical wave vector
{\bf q}$_0$, where the dominant instability occurs.
The localized nature of the anomaly in q-space is a signature of its 
electronic origin, as will be discussed in detail in the next section.

\subsection {Origin of the anomaly}

Screening due to electron-phonon coupling involving electronic states
near the Fermi level in metals can give rise to anomalous dips in the
phonon dispersion, called Kohn anomalies.  The occurrence of these
anomalies depends mainly on the geometry of the Fermi surface, as well
as on the {\bf q}-dependence of electron-phonon matrix elements.  If
the Fermi surface has flat portions with nesting vector {\bf q}$_0$,
there generally will be very strong screening of the potential
perturbation due to atomic displacements at that wavevector, leading
to a pronounced softening highly localized in q-space. The effect of
Fermi surface geometry can be quantified by calculation of the
generalized susceptibility, as in Ref. \onlinecite{harmon2002}.

\begin{figure}
\vbox{
 \centerline{\psfig{figure=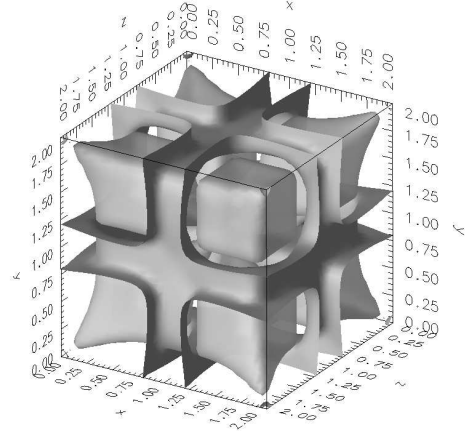,width=6cm,angle=0}}
\vspace{0.4cm}
}
\caption{
Fermi surface of minority spin bands for cubic Ni$_2$MnGa.}
\label{fig:FermiSurfCubic}
\end{figure}

In cubic Ni$_2$MnGa, the Fermi level crosses both minority and
majority spin bands.  Of the two Fermi surfaces, only the one for the
minority spin bands, plotted in figure~\ref{fig:FermiSurfCubic}, shows
obvious nesting features. In particular, a large fraction of 
the opposite sides of the
``pipes,'' running along the faces of the cube and crossing at their
center, are nested by wavevectors of the form 
($\zeta_0, k, 0$)${2 \pi \over a}$
with $\zeta_0$ = 0.34 and $0 \leq k \leq  0.34$ and the
cubic-symmetry-related equivalent wavectors. Along [110], this
corresponds to the critical wave vector of the anomaly. It should be
noted that this differs from the position of the nesting wavevector
and peak in generalized susceptibility along [110] found in Refs.
\onlinecite{harmon2002} and \onlinecite{naumov}. This difference might
be attributable to the differences in method, in particular, the
difference in choice of density functional. [110] appears to be a
direction for which the anomaly is strongest. For example, although
there is nesting along $[$100$]$ as well, there is no sign of an
anomaly in the phonon dispersion.  This can be attributed to the
vanishing of the relevant electron-phonon matrix elements.

The strength of a Kohn anomaly is expected to be very sensitive to
electronic temperature, with an increasing temperature reducing the
sharpness of the Fermi surface and thus weakening the anomaly.  
In our calculations,
the sensitivity to electronic temperature can be directly
investigated, as the smearing parameter $\sigma$ plays the role of a
fictitious electronic temperature.  In table~\ref{tab:sigma3conv} we
show the dependence on $\sigma$ of those modes at the critical wave
vector, {\bf q}$_0$, that have the same symmetry as the soft TA$_2$
branch (i.e. the three modes of $\Sigma_3$ symmetry).  The anomalous
TA$_2$ mode, $\Sigma_3(1)$, is much more sensitive to the fictitious
electronic temperature than the other two modes.  All the frequencies
are well converged for $\sigma$=0.01 Ry.  The dependence of the phonon
anomaly upon the electronic temperature is also shown in
figure~\ref{fig:phcubic}, where the dispersion of the anomalous TA$_2$
branch computed with $\sigma$=0.03~Ry (dashed line) is compared with
the fully converged calculation for $\sigma$=0.01~Ry (solid line).
Only the modes with wavevectors close to the anomaly,
($\zeta_0-0.16)<\zeta<(\zeta_0+0.16$), are affected by the change in
the electronic temperature. The anomaly is smoothed out as the
fictitious electronic temperature is increased.  Thus, we conclude
that the computed dip is indeed a Kohn anomaly.

\subsection{Elastic constants}

The computed elastic constants for the ferromagnetic cubic structure
are reported in table~\ref{tab:elastic}.  In a first-principles
framework, the standard approach for obtaining elastic constants is to
compute the stress tensor for a selected set of small strains.  To
obtain convergence at the level of a few GPa, we have used very dense
k-point meshes (up to (14 14 14), corresponding to 10976 points in the
BZ).  To compute C$_{11}$ and C$_{12}$ the cubic crystal is distorted
by the tetragonal deformation: $\epsilon_{zz}$=$\epsilon$,
$\epsilon_{xx}$=$\epsilon_{yy}$=0. For small deformations the stress,
$\sigma_{ij}$, is linear with $\epsilon$:
$\sigma_{xx}$=$\sigma_{yy}$=C$_{12}\epsilon$, and
$\sigma_{zz}$=C$_{11}\epsilon$. The values so obtained for the elastic
constants C$_{11}$, C$_{12}$, and the shear modulus 
C$^\prime$=(C$_{11}-$C$_{12}$)/2 are
indicated as ``theory 1'' in table~\ref{tab:elastic}.

The small magnitude of $C^\prime$ requires additional attention.  We
performed an independent computation by considering the tetragonal
deformation $\epsilon_{xx}$=$\epsilon_{yy}$=$\epsilon$,
$\epsilon_{zz}$=$-$2$\epsilon$.  For small $\epsilon$ we have that
$\sigma_{xx}$=$\sigma_{yy}$=2$C^\prime\epsilon$ and
$\sigma_{zz}$=$-4$C$^\prime\epsilon$. This direct calculation gives
C$^\prime$=(2$\pm$2) GPa, in good agreement with the value obtained as
a difference of C$_{11}$ and C$_{12}$ in theory 1.  The near-zero
value of $C^\prime$ implies that there is almost no energy cost for a
small tetragonal shear distortion of the type
$2\epsilon_{xx}$=2$\epsilon_{yy}$=$-\epsilon_{zz}$.  Evidence of this
marginal elastic instability towards small volume-preserving
tetragonal distortions was also found in Ref.\onlinecite{vitaly01},
where it was shown that the energy surface as a function of $c/a$ is
remarkably flat; specifically, the change in energy associated with
varying the $c/a$ ratio in the range of values between 0.97 and 1.01
is almost zero to within numerical accuracy.

In table~\ref{tab:elastic} (theory 2) we also show the values of the
elastic constants estimated from the slope of the long wavelength
acoustic modes along the [110] direction. The elastic constants
C$_{44}$, C$^\prime$, and C$_L$=(C$_{11}$+C$_{12}$+2C$_{44}$)/2
correspond to the TA$_1$, TA$_2$, and LA modes, respectively. These
determine the values of $C_{11}$ and $C_{12}$ given in the table.  The
evaluation of the elastic constants from the phonon dispersion is less
accurate, especially for C$^\prime$ where the corresponding TA$_2$
branch deviates from a linear behavior already at very small q.

The elastic constants in the high-temperature cubic phase have
been experimentally determined between room temperature and the 
pre-martensitic structural phase transition at T = 260 K
by measuring the velocity of ultrasonic waves.~\cite{Worgull96,Manosa97}
These experiments show that the elastic constants 
$C_L$ and $C_{11}$ are almost temperature independent.
In contrast, the transverse elastic constants $C_{44}$ and 
$C^\prime$ exhibit an anomalous behavior with cooling,
softening as the pre-martensitic phase transition temperature is approached.
While the softening is small for $C_{44}$, it is more dramatic 
for $C^\prime$, which decreases by 60\% from room temperature to the 
transition.
We find good agreement between 
theory and room temperature measurements for $C_L$ and $C_{11}$, which are
not strongly modified by finite temperature.
The pronounced softening of $C^\prime$ is also consistent with the
very small zero-temperature theoretical value, $C^\prime\approx0$.

\section{Tetragonal N\lowercase{i}$_2$M\lowercase{n}G\lowercase{a}}

The identification of unstable phonons at 
{\bf q}$_0$=${2\pi\over a}$($\zeta_0$, $\zeta_0$, 0), 
with $\zeta_0=0.34$, in the cubic
Heusler structure leads naturally to an understanding of the
transition to the premartensitic cubic-modulated phase with 
decreasing temperature.
To explore the subsequent transition to the low-temperature
martensitic phase with $c/a$ = 0.94 and modulation wavevector 
$\zeta_0=0.43$, we extended the calculations of phonon dispersion to
Heusler structures with uniform volume-preserving tetragonal strains
ranging from $c/a$ = 0.88 to $c/a$ = 1.06, with particular attention 
to the TA modes along the $<$110$>$ directions.

\subsection{Tetragonal structure with $c/a$=0.94}

\begin{figure}
\vbox{
\centerline{
 \psfig{figure=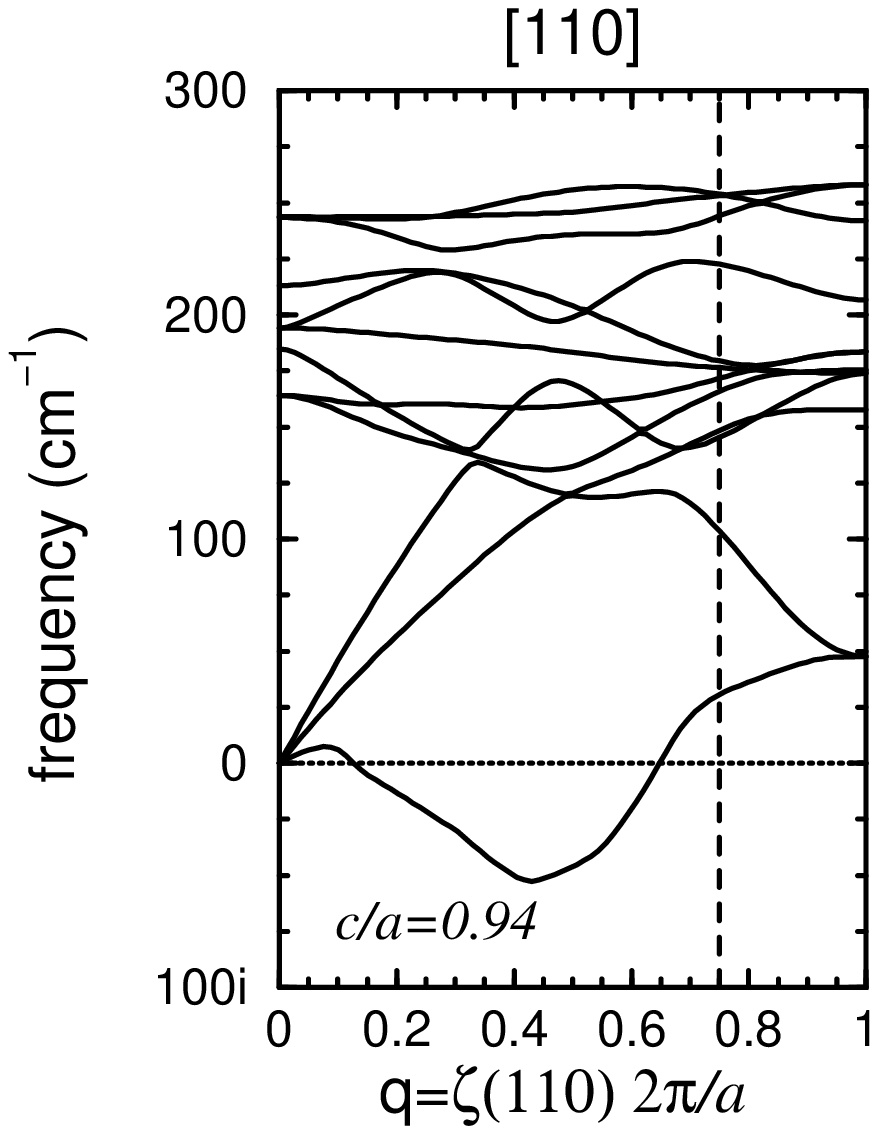,height=6cm,angle=0}
 \psfig{figure=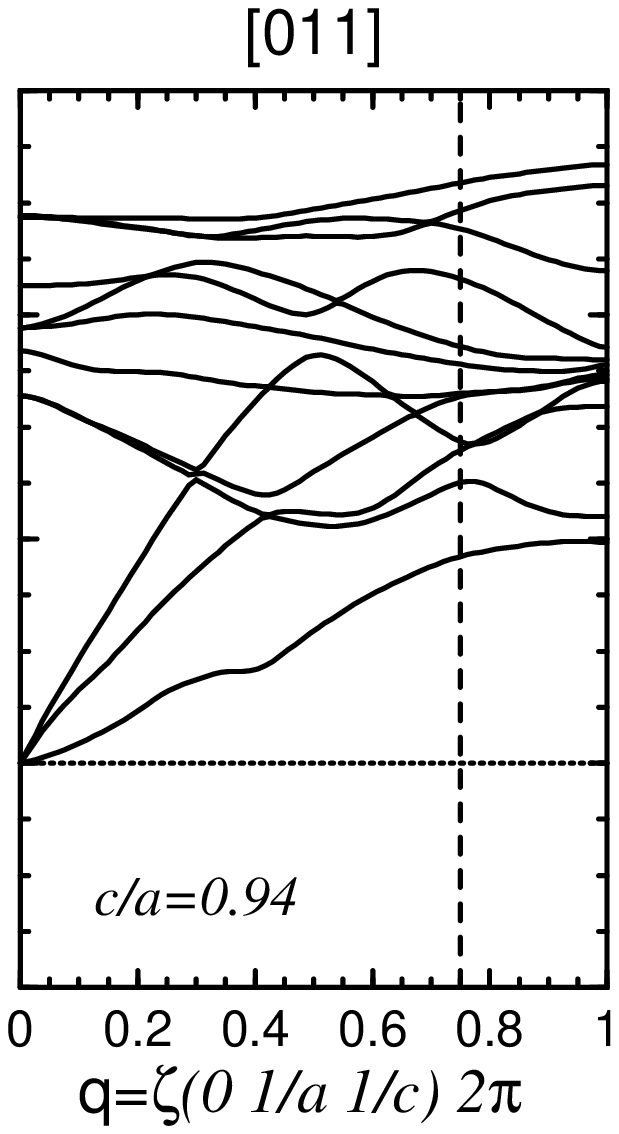,height=6cm,angle=0}
}
}
\caption{Ab initio [110] and [011] phonon dispersions for ferromagnetic 
Ni$_2$MnGa in the tetragonal structure ($c/a$=0.94, $a$=11.26 a.u.)
}
\label{fig:phtetragonal}
\end{figure}

We first consider in detail the volume-preserving tetragonal distortion with 
$c/a$=0.94,
corresponding to the low temperature martensitic phase.
The phonon dispersions computed along the inequivalent [110] and [011] 
directions of the tetragonal BZ are shown in figure~\ref{fig:phtetragonal}.
The only mode substantially affected by the tetragonal distortion is the
anomalous TA$_2$ mode. 
We find an overall softening (hardening) of the TA$_2$ branch in the [110] 
([011]) direction. 
The anomaly is particularly affected by the tetragonal 
distortion.
Along the $<$110$>$ directions, which are
perpendicular to the $c$ axis, the phonon anomaly
is more pronounced and is shifted to a larger q-vector ($\zeta=0.43$) 
than in the cubic structure. 
This wavevector is in excellent agreement with the long period 
modulation observed experimentally in the tetragonal martensitic phase. 
In contrast, along the $<$011$>$ directions, the anomaly has almost completely
disappeared; the entire branch is stable and there is only a very small wiggle
in the TA$_2$ branch. Thus, the entropy from low-frequency phonons should be
roughly a factor of three less than in the premartensitic phase, which may 
explain why the premartensitic phase is more favorable at higher temperatures.

\begin{figure}
\vbox{
 \centerline{\psfig{figure=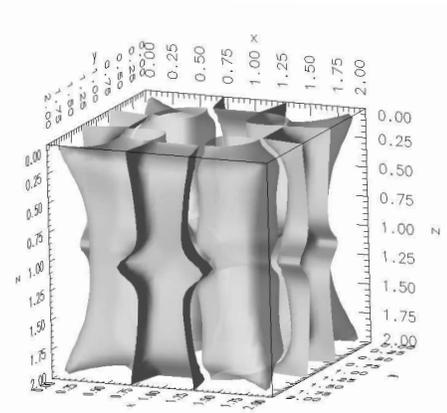,width=6cm,angle=-1}}
\vspace{.5cm}
}
\caption{Fermi surface of minority spin bands for tetragonal 
Ni$_2$MnGa ($c/a$=0.94).}
\label{fig:FermiSurfTetr}
\end{figure}

As in the cubic phase, the anomaly can be associated with features of the 
Fermi surface.
In figure~\ref{fig:FermiSurfTetr} we show the minority-spin Fermi surface
computed for Ni$_2$MnGa in the tetragonal structure.
It is related to the Fermi surface of the cubic structure with 
two major differences: (i) there are no flat surfaces perpendicular to
the (001) direction, and
(ii) the ``pipes'' running along the 
faces of the tetragonal BZ are wider than the corresponding features 
in the cubic Fermi surface.
As a consequence of these changes, 
induced by the tetragonal distortion, there
is significant nesting only in the $<$110$>$ directions perpendicular
to the $c$ axis and the edges of the ``pipes'' are nested by the
larger ($\zeta$ $\zeta$ 0) critical wave-vector, $\zeta=0.43$.
This explains why in the tetragonal structure there are pronounced Kohn 
anomalies only in the $<$110$>$ directions, at a larger wave vector 
than in the cubic structure.

\subsection{Evolution of phonon anomaly with tetragonal deformation}

To understand better the lattice instabilities and related structural
energetics of Ni$_2$MnGa, we studied the evolution of the TA$_2$
phonon anomaly with uniform volume-preserving tetragonal strain.

Studying the phonon dispersions of the cubic Heusler structure
($c/a=1$) and its tetragonal distortion with $c/a=0.94$, we have
identified the existence of the unstable mode at $\zeta=0.34$ and
$\zeta=0.43$, respectively, that explain the transitions to the
premartensitic and martensitic phases with decreasing temperature. The
fact that for the two tetragonal strains the soft mode is always the
TA$_2$ mode but with a different wave vector suggests that the
wavevector of the soft mode and the tetragonal strain are coupled.  
We therefore studied in more detail the dependence of $\zeta$ upon
tetragonal strain.

\begin{figure}
\vbox{
 \centerline{\psfig{figure=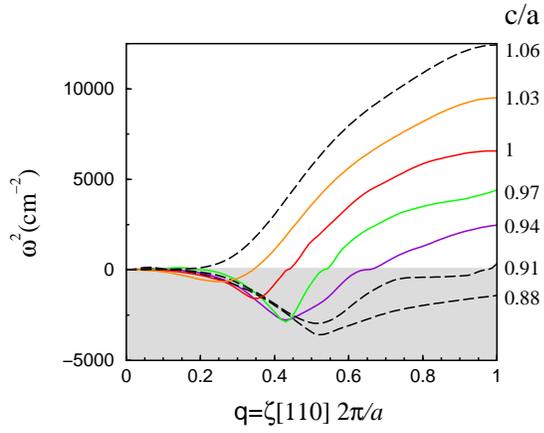,width=7cm,angle=0}}
}
\caption{(color online only). 
Dispersion of the squared frequencies, for the anomalous TA$_2$ phonon 
branch, computed for different
volume-preserving tetragonal distortions, with $c/a$ ranging from 
0.88 to 1.06.
}
\label{fig:anomalytetragonal}
\end{figure}

The dispersion of the TA$_2$ branch along the [110] direction is shown
in figure~\ref{fig:anomalytetragonal} for several values of the $c/a$
ratio.  An overall softening of the TA$_2$ branch is observed with
decreasing $c/a$ ratio.  With compressive strain ($c/a<1$) the anomaly
occurs at a larger $\zeta_0$ than in the cubic structure, and becomes
broader and more pronounced.  For $c/a\lesssim 0.91$ the entire branch
is unstable.  With tensile strain ($c/a>1$) the anomaly shifts to a
smaller $\zeta_0$ and becomes less pronounced. The lattice instability
is completely eliminated for $c/a\gtrsim 1.06$. This is in agreement
with the experimental observation of no superimposed modulations in a
stress induced tetragonal phase with $c/a$ = 1.18.~\cite{Martynov95}

In figure~\ref{fig:anomalytetragonal}, we have shown the TA$_2$ branch
for several different values of $c/a$. For each of these values of
$c/a$, we can find the value of the wavevector $\zeta_0$ for which the
imaginary frequency of the TA$_2$ branch has its minimum. This gives
us the dependence of the modulation wavevector upon strain and is
shown in figure~\ref{fig:anomalytetragonal}. Three regimes can be
identified: (i) $c/a\lesssim 0.91$, where the entire TA$_2$ branch is
unstable and there is no unique minimum; (ii) $0.9<c/a<1.06$, where
the TA$_2$ branch displays a well-defined dynamical instability
localized at the wavevector $\zeta_0$ that depends on $c/a$; and (iii)
$c/a\gtrsim 1.06$, where the lattice instability is completely
eliminated and the TA$_2$ branch is stable over the whole BZ, so that
no minimum is defined. Only in regime (ii) can a modulated structure be
expected to occur. 

\begin{figure}
\vbox{
 \centerline{\psfig{figure=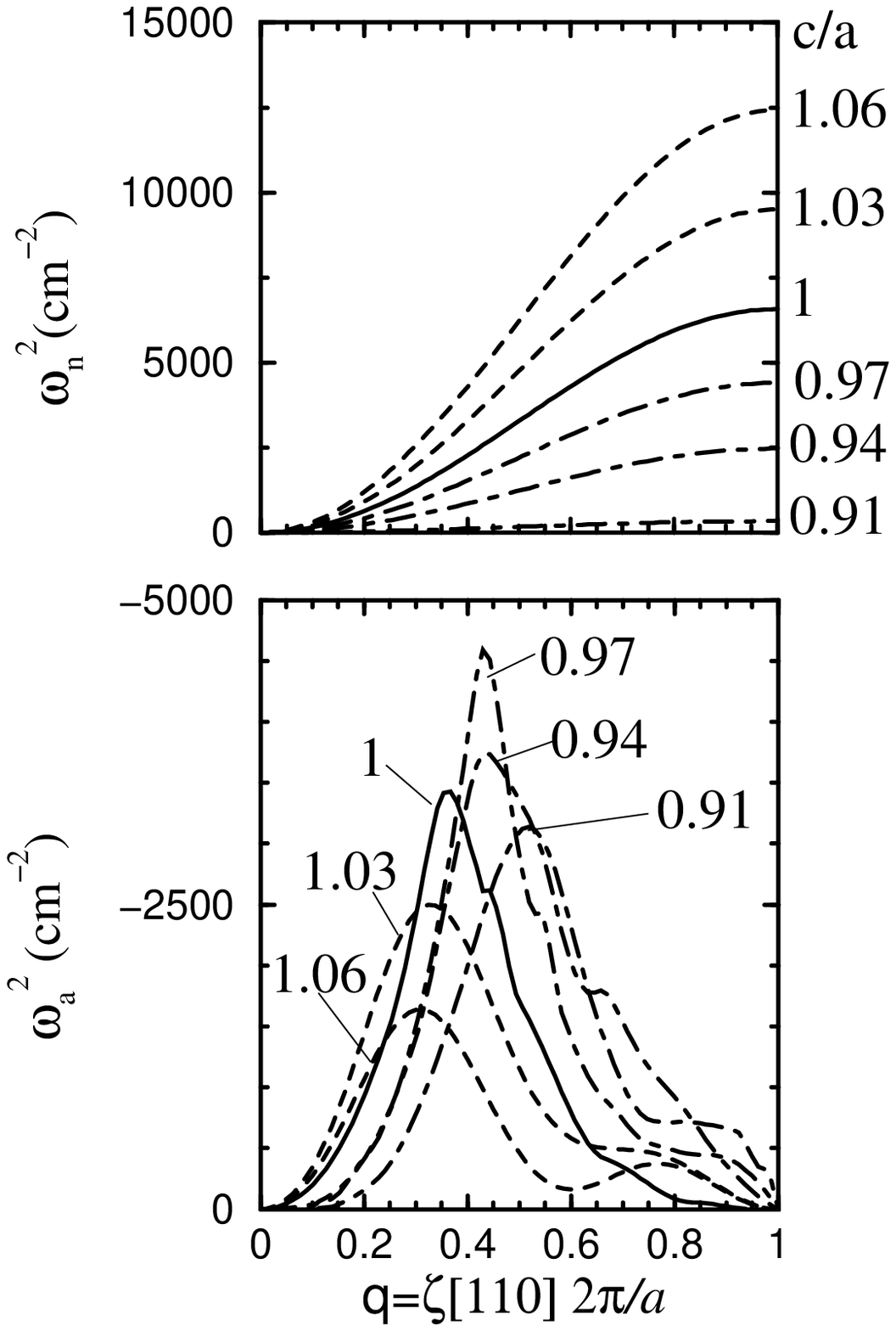,width=6cm,angle=0}}
}
\caption{
Decomposition of $\omega^2$ into ``normal'' ($\omega^2_n$) and anomalous
($\omega^2_a$) parts for different volume-preserving tetragonal distortions, 
with $c/a$ varying from 0.91 to 1.06. Dashed lines are for $c/a>1$, 
dot-dashed lines are for $c/a<1$ and solid lines are for $c/a=1$.
}
\label{fig:anomalydecomposition}
\end{figure}

In order to quantify these trends, in figure~\ref{fig:anomalydecomposition} 
we decompose the phonon frequency $\omega^2$ into a 
short-range ``normal'' part $\omega^2_n$
and a long-range ``anomalous'' part $\omega^2_a$: 
$\omega^2(\zeta)$=$\omega^2_n(\zeta)+\omega^2_a(\zeta)$.
The ``normal'' contribution, defined as 
$\omega^2_n(\zeta)$=$A[1-cos(\zeta\pi)]$, depends only upon short-range 
interatomic force constants and is connected to the local chemistry 
of the crystal. It corresponds to the simple model for the 
transverse [1$\overline1$0] vibrations of 
the [110] planes when they interact only with a first-neighbor
interplanar force constant: K=-A$m$/2, where $m$ is the mass associated
with each plane.
The long-range
``anomalous'' contribution $\omega^2_a$ depends on electronic 
screening effects that can be strongly {\bf q} dependent.
To fit the constant A in the definition of $\omega^2_n(\zeta)$, we assume that 
for $\zeta$=1 the anomalous contribution is zero.

The short-range contribution 
softens with decreasing $c/a$ ratio, causing the overall 
softening observed for the TA$_2$ branch. 
Decreasing $c/a$ corresponds to a volume-preserving tetragonal distortion with 
a shorter $c$ and larger $a$. Therefore smaller $c/a$ values correspond to a 
larger [110] interplanar distance which leads to a weaker interplanar 
force constant and therefore a softer $\omega^2_n$.

The anomalous contribution $\omega^2_a$ is negative, large, and peaked at a 
critical wave vector $\zeta_0$, indicating a strong screening due to the 
electon-phonon coupling and Fermi surface nesting. 
Both $\zeta_0$ and the intensity of the anomaly are
strongly dependent upon strain. The value of $\zeta_0$ increases with 
decreasing $c/a$, the intensity reaching its maximum at $c/a$=0.97.

\section {Summary and conclusions}

Our ab initio study of the phonon dispersion and lattice instabilities 
in cubic and tetragonal Ni$_2$MnGa can be summarized as follows.

First, the parent cubic phase exhibits a marginal elastic instability,
C$^\prime\approx0$ , meaning that there is almost no energy cost for a
small uniform tetragonal distortion.

Second, for $0.91<c/a<1.06$ a Kohn anomaly is present due to
electron-phonon coupling and Fermi surface nesting.  It develops into
a deep minimum corresponding to a localized dynamical instability
whose wave vector $\zeta_0$ is related in a one-to-one fashion to the
$c/a$ ratio. Therefore, tetragonal structures with $c/a$ in this range
are unstable towards a particular transverse (110)-shuffling
modulation with a specific wavevector $\zeta_0$, the one-to-one
relationship of the $c/a$ ratio and the modulation period being due to
the behavior of the topology of the Fermi surface under tetragonal
distortions.

Consequently, the energy of the cubic structure may be lowered by any
one of a family of deformations, $\{\epsilon, {\bf
u}(\zeta_\epsilon)\}$, consisting of a tetragonal strain, $\epsilon$,
and a superimposed (110)-shuffling modulation, ${\bf
u}(\zeta_\epsilon)$, whose periodicity, $\zeta$, is directly related
in a one-to-one fashion to the $c/a$ ratio.

The near-vanishing of C$^\prime$ together with the fact that the
energy surface as a function of $c/a$ is remarkably flat permits a
modulated ground state with $c/a$ significantly different from 1.
Indeed, the computed relation between $c/a$ and the modulation
wavevector is in excellent agreement with structural data on the
premartensitic ($c/a$ = 1) and martensitic ($c/a$ = 0.94) phases of
Ni$_2$MnGa, and also with the unmodulated stress-induced phase ($c/a$
= 1.18).

The next step is to carry out total energy calculations of the complex
modulated structures in order to determine the equilibrium amplitude
and phase of the lattice modulation and the associated energy
gain. The energy could then be expanded in terms of symmetry
invariants of the relevant structural degrees of freedom (strain and
modulation) and determine the coefficients of each term of the
expansion using ab initio calculations. In this way one could develop
an ab-initio based finite-temperature statistical model of the phase
transition in Ni$_2$MnGa.

The three phases of Ni$_2$MnGa observed with increasing temperature
can be understood as follows. At low temperatures, a minimum-energy
modulated tetragonal structure is expected, consistent with the
observed $c/a$ = 0.94 phase.  As temperature increases, the favorable
entropy arising from the larger phase space for the low-frequency
phonons in the cubic structures leads to a phase transition to the
cubic-modulated structure, with a modulation wavevector shifted down
from that in the tetragonal phase in accordance with our
results. Finally, the frequencies of the unstable phonons associated
with the modulation are anharmonically renormalized at the highest
temperatures and the observed structure is the cubic Heusler
structure, with large fluctuating local distortions.

In conclusion, we have provided a first-principles microscopic explanation
for the origin of the physically interesting and technologically important
martensitic phase transitions in the ferromagnetic shape-memory Heusler alloy
Ni$_2$MnGa.

\vspace{0.7cm}
Calculations in this work have been done using the PWscf package.~\cite{PWSCF}
We thank Morrel Cohen, Richard James and Xiangyang Huang for valuable 
discussions, 
and Vitaly Godlevsky for his help in the early stages of this work.

\begin{table}
\caption{
Equilibrium lattice parameter, $a_0$, bulk modulus,
B$_0$, and magnetic moment, $\mu_0$, for Ni$_2$MnGa in the L2$_1$
Heusler structure. Calculations using the spin-polarized GGA ($\sigma$-GGA),
are compared with experimental data and with the results obtained using LSDA.
}
\begin{tabular}{l c c c}
             & $a_0$(a.u.) &  B$_0$(Mbar) & $\mu_0$($\mu_{\rm B}$)  \\
\hline 
$\sigma$-GGA &  11.03      &  155         &  4.27            \\ 
LSDA         &  10.74      &  202         &  3.92            \\ 
FLAPW-$\sigma$-GGA$^c$&
                10.98      &  156         &  4.09                  \\
Exp.         &  11.01$^a$  &  146$^b$     &  4.17$^a$              \\ 
\end{tabular}
$^a$Ref.\onlinecite{Webster69}\\
$^b$Ref.\onlinecite{Worgull96}\\
$^c$Ref.\onlinecite{Ayuela99}\\
\label{tab:structure}
\end{table}
\begin{table}
\caption{Phonon frequencies (in cm$^{-1}$) computed at the zone-boundary 
point X. Results obtained using different approximations for the  
exchange and correlation energy, $\sigma$-GGA and LSDA, are compared
with the experiment.}
\begin{tabular}{l c c c c }
   & $\sigma$-GGA & LSDA & LSDA                  & Exp. \\ 
$a_0$  
&  $a_0$($\sigma$-GGA)& $a_0$(LSDA) & $a_0$($\sigma$-GGA)   &      \\
\hline
X$_5^\prime$ &
       82         & 76   &  70                   & 87   \\
X$_1^\prime$ 
   &   160        & 179  & 151                   &      \\
X$_5$ 
   &   174        & 191  & 161                   &      \\
X$_5^\prime$
   &   176        & 195  & 166                   &      \\
X$_4^\prime$
   &   178        & 185  & 159                   &      \\
X$_1$
   &   191        & 203  & 168                   &      \\
X$_1^\prime$
   &   227        & 249  & 212                   &      \\
X$_5^\prime$ 
   &   259        & 286  & 233                   &      \\
\end{tabular}
\label{tab:Xfreq}
\end{table}
\begin{table}
\caption{
Dependence upon the smearing parameter, $\sigma$, 
of the frequency of the three $\Sigma_3$-modes
at the critical wavevector, {\bf q$_0$}.
The $\Sigma_3(1)$ mode is the anomalous TA$_2$ mode, 
the other two, $\Sigma_3(2)$ and $\Sigma_3(3)$, are
higher frequency optical modes.
The {\bf k}-point mesh needed to achieve 
convergence for each of the values of $\sigma$ is also reported.
Frequencies are in cm$^{-1}$.
}
\begin{tabular}{l c c c c}
$\sigma$(Ry)  & {\bf k}-mesh  & $\Sigma_3(1)$ & $\Sigma_3(2)$ & $\Sigma_3(3)$ \\
\hline 
0.03      &(6  6  6)   &    8      &       195     &     244      \\
0.01      &(10 10 10)  &   39$i$  &       195     &     242      \\
0.005     &(12 12 12)  &   39$i$  &       195     &     242      \\
\end{tabular}
\label{tab:sigma3conv}
\end{table}

\begin{table}
\caption{ Elastic constants, in GPa, for ferromagnetic Ni$_2$MnGa in the
cubic Heusler structure. 
Theory 1, elastic constants from stress calculation under strain. 
Theory 2, elastic constants estimated 
from the slope of the [$\zeta \zeta 0$] acoustic branches close to $\Gamma$: 
C$^\prime$, C$_{44}$, and C$_{\rm L}$ correspond to the TA$_2$, TA$_1$, and 
LA modes, respectively. 
Room temperature measurements are shown for comparison.
}
\begin{tabular}{l c c c c c}
	& C$_{11}$ &  C$_{12}$ &  C$^\prime$  & C$_{44}$  &  C$_{\rm L}$ \\
\hline 
theory 1 &153$\pm$2 & 148$\pm$2 &  2.5$\pm$2    &           &              \\
theory 2 & 138$\pm$9   & 143$\pm$9 &  $-$2.5$\pm$5 &100$\pm$5 &  240$\pm$5 \\
Exp.$^a$&152       &   143     &     4.5      & 103       &    250       \\
Exp.$^b$&136$\pm$3 &    --     &    22$\pm$2  & 102$\pm$3 &    222$\pm$9 \\
\end{tabular}
$^a$Ref.\onlinecite{Worgull96}\\
$^b$Ref.\onlinecite{Manosa97}\\
\label{tab:elastic}
\end{table}

\onecolumn

\twocolumn


\begin{figure}
\vbox{
 \centerline{\psfig{figure=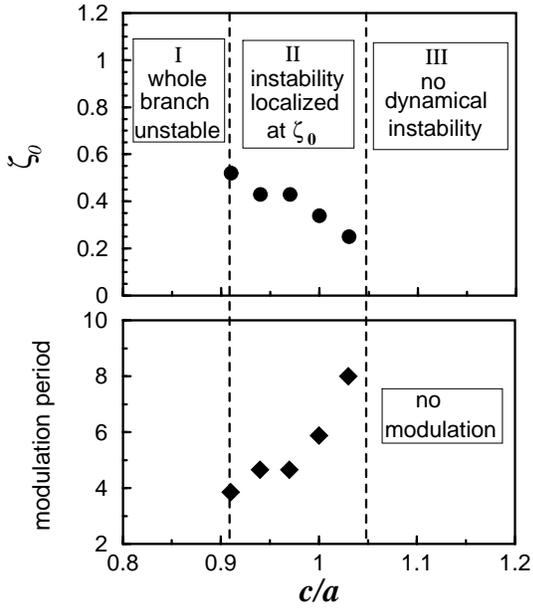,width=7cm,angle=0}}
}
\caption{
Dependence upon $c/a$ of the modulation wavevector $\zeta_0$ and the 
related modulation period (in unit of (110) interplanar distances). 
}
\label{fig:zetavscova}
\end{figure}

\begin{references}
\bibitem{Nips} The details for the Ni pseudopotential can be found 
               in Ref.\protect\onlinecite{DalCorso00}.
\bibitem{Mnps} We generated the pseudopotential for Mn, using as reference the 
               all-electron configuration 3$d^5$, 4$s^2$. The core radii are
               (in a.u.): 4$s$ (2.0, 2.2), 4$p$ (2.2, 2.4), 3$d$ (1.6, 2.2).
               The all electron potential pseudized inside r$_c$=1.8 has been 
               taken as a local potential. Two values of the core radii 
               indicate a channel which has been pseudized with the ultrasoft 
               scheme. In such a case the first value is the norm-conserving
               core radius and the second is the ultrasoft one.
\end{references}
\end{document}